\documentclass[times, longbib, modern]{aastex701}

\usepackage{amsmath}

\usepackage[T1]{fontenc}

\usepackage{etoolbox}
\makeatletter
\patchcmd{\NAT@citex}
  {\@citea\NAT@hyper@{%
     \NAT@nmfmt{\NAT@nm}%
     \hyper@natlinkbreak{\NAT@aysep\NAT@spacechar}{\@citeb\@extra@b@citeb}%
     \NAT@date}}
  {\@citea\NAT@nmfmt{\NAT@nm}%
   \NAT@aysep\NAT@spacechar\NAT@hyper@{\NAT@date}}{}{}
\patchcmd{\NAT@citex}
  {\@citea\NAT@hyper@{%
     \NAT@nmfmt{\NAT@nm}%
     \hyper@natlinkbreak{\NAT@spacechar\NAT@@open\if*#1*\else#1\NAT@spacechar\fi}%
       {\@citeb\@extra@b@citeb}%
     \NAT@date}}
  {\@citea\NAT@nmfmt{\NAT@nm}%
   \NAT@spacechar\NAT@@open\if*#1*\else#1\NAT@spacechar\fi\NAT@hyper@{\NAT@date}}
  {}{}
\makeatother

\usepackage{natbib}
\newcommand\affilcfa{Center for Astrophysics | Harvard \& Smithsonian, 60 Garden Street,
  Cambridge, MA 02138, USA}

\newcommand\apx{\ensuremath{\sim}}
\let\b=\textbf
\newcommand\citeeg[1]{\citep[e.g.,][]{#1}}
\let\i=\textit

\newcommand\fnurl[1]{\footnote{\url{#1}}}

\begin{document}

\title{AI and The Enclosure of the Scientific Commons}
\author[0000-0003-3734-3587]{P. K. G. Williams}
\email{pwilliams@cfa.harvard.edu}
\affiliation{\affilcfa}
\shorttitle{AI and The Enclosure of the Scientific Commons}
\shortauthors{Williams}

\begin{abstract}
  \noindent The scientific literature can be viewed as an ``information
  commons:'' an intangible resource that ``no one owns and is shared by
  everyone.'' There are well-established economic and political frameworks for
  understanding such commons, emerging from the history of agricultural common
  lands found in places like preindustrial England. This article provides a
  primer on the history and these frameworks, and considers a key difference
  between the information commons and the fields of yore: while information
  commons are robust and immensely valuable --- some can be said to rank among
  humanity's greatest achievements --- the English agricultural commons no
  longer meaningfully exist. The latter are gone not because of the well-known
  ``tragedy of the commons,'' but rather a phenomenon known as the enclosure
  movement ($\sim$1600--1800) that transformed the common lands into private
  property. Many observers, then and now, judge enclosure to have been theft at
  a societal scale. This article argues that artificial intelligence (AI)
  provides precisely the tools needed to enable a previously-infeasible
  enclosure of the information commons. Although AI-enabled enclosure does not
  operate identically to physical enclosure, it shares the key features of
  turning the commons into private wealth and undermining the social systems
  that support them (via ``slop spam''). The concepts presented in this article
  can provide a scholarly basis for interrogating the foreboding feeling that AI
  may represent not just a smash-and-grab, but an attack upon a whole way of
  life.
\end{abstract}

\section{Introduction}

\noindent Although the artificial intelligence (AI) industry is undoubtedly
fueled by a large amount of hype and some further amount of deception, it is no
longer tenable to argue that it's \i{all} smoke and mirrors. AI tools can do
useful and interesting things, and in at least some situations they enable
technical projects that would heretofore have been unthinkable. This can be cast
as a dramatic reduction in the costs of certain activities.

Under genuine economic pressure, it would require a person to have an extreme
ideological commitment to practice ``AI abstinence'' and refuse its use. Some
people do have such extreme commitments, though. Recalcitrants in the face of
the current AI mania have been called Luddites, in reference to the skilled
English textile workers who violently opposed the mechanization of their
industry, intimidating factory owners and destroying machinery
\citep{hobsbawm52}. The label is generally meant to be derogatory, but it's
indisputable that the Luddites are victims of ``the enormous condescension of
posterity'' \citep{thompson63}. Luddites understood correctly that their
relatively high-status way of life was under threat, and it hardly seems fair to
fault them for failing to overcome the combination of inexorable capitalist
logic and violent state opposition.

This article seeks to provide a framework for understanding why some of us may
be feeling a bit Luddite these days. It highlights a related but separate
phenomenon from the time of English industrialization: the \b{enclosure
movement}. In many ways this article is a primer on the concept of \b{the
commons}, both historical and present-day, aimed at an audience of scientists
and engineers who may not have been exposed to the related history and theory
before.

\autoref{s.commons} introduces the historical agricultural commons of England
and their enclosure. \autoref{s.modern} explains how the concept of the commons
remains relevant to the modern world, even though its namesake pastures no
longer meaningfully exist. The practice of science and scholarship can be seen
as an example of a phenomenon known as \b{commons-based peer production},
described in \autoref{s.peer}. \autoref{s.ai} considers the current AI moment in
light of these concepts, and \autoref{s.newenc} argues that it can be viewed as
the potential beginning of a new enclosure movement. \autoref{s.conc} offers
some concluding remarks.

For brevity, this article simply refers to ``AI'' and ``AI companies'' without
routinely identifying specific model types or corporations. The discussion is
predominantly relevant to the generative AI and large language models (LLMs)
like ChatGPT (OpenAI) or Claude (Anthropic), but does not depend on their
details. This article also aims to avoid reliance on extrapolations about the
future capabilities of AI systems, the evolution of the AI industry, and
potential impact of these on the rest of society. Such extrapolations are worth
exploring, but make for unsteady ground upon which to conduct debate
\citep{es03}. Separately, this article omits discussion of the impacts of AI
adoption on a variety of topics such as the environment, public health, and
equity. These considerations are important, but beyond the scope of the present
discussion.

\section{The Historical Commons}
\label{s.commons}

\noindent Commons, in the sense of communally-owned land or other natural
resources, have existed since antiquity and continue to exist in the present day
\citeeg{wall14, ostrom90}. This article's analysis focuses on what are perhaps
the most famous, and surely the best-studied, commons: the agricultural lands of
feudal and preindustrial England \citeeg{neeson93, winchester22}. In that
system, the common land was understood to be owned by a local lord, but members
of the local community could have rights to use the commons for activities such
as small-scale agriculture, grazing, and the collection of firewood. Hence the
term ``commoner'': a person with commons rights whose livelihood therefore
depended upon the commons.

From a modern perspective, the commons can feel foreign because economically
useful lands are now generally managed under exclusive, private-property
regimes, rather than as a shared resource. But this should not be taken to mean
that the commons were a free-for-all: the actual operation of the preindustrial
commons could be quite sophisticated \citep{thompson91} and is the topic of
significant historical study that is far too broad to usefully summarize here.

\subsection{The (Real) Tragedy of the Commons}

\noindent For many readers, an association game with the word ``commons'' will
instantly yield the word ``tragedy.''

The widely-known concept of ``the tragedy of the commons'' was articulated by
\citet{hardin68}, in an article of the same name. The basic argument, which the
author himself acknowledges is not wholly novel, is that commons are not
sustainable: each participant is motivated to extract as much individual value
out of the commons as possible, inevitably leading to over-use, exhaustion, and
collapse.

More nuanced work makes clear, however, that the tragedy of the commons is at
best an oversimplification, and arguably a piece of propaganda masquerading as
research. (The primary subject of Hardin's article is ``overpopulation'' and his
proposed solution is essentially eugenics --- one of the section titles is
``Freedom to Breed is Intolerable.'' Hardin has been classified as a white
nationalist by the Southern Poverty Law
Center\fnurl{https://www.splcenter.org/resources/extremist-files/garrett-hardin/}.)
Out of the copious literature exploring this topic, one of the most famous works
is that of \citet{ostrom90}. She analyzes commons theoretically and identifies a
number of criteria that can enable their sustainability, as well as real-world
examples of present-day, successful commons. Approximately, commons need some
form of boundary policing, so that not just \i{anyone} can come in and do
whatever they want, as well as internal governance structures.

That may be the case, but it remains true that the commons of preindustrial
England no longer meaningfully exist: present-day common lands are
predominatedly used for recreation and conservation, rather than supporting
livelihoods \citep{winchester22}. If we are to identify a real, singular tragedy
of the historical commons, it is not that they collapsed; it is that they were
stolen.

\subsection{The Enclosure Movement}
\label{s.enclosure}

\noindent The destruction of the English commons was accomplished as part of what
is known as the enclosure movement. \citet{fairlie09} offers the non-specialist
an entry point to some of its history. Over a span of centuries, ranging
approximately from the 1600's to the 1800's, English lands, including the
commons, were brought under a private property regime, with newly-codified
boundaries enforced by physical barriers such as hedges and ditches. The process
was largely driven by wealthy landowners, who typically became formal owners of
the majority of the enclosed land. It generally resulted in commoners losing
their traditional land access rights, leading to social unrest and an uprooted
populace that gradually transformed into the factory workforce that powered the
Industrial Revolution.

The enclosure movement has been politically contested for the better part of a
millennium and is the subject of intense historical study. Enclosure is often
interpreted, justifiably, as a confrontation between capitalism and traditional
lifestyles. The arguments for enclosure generally centered on economic rewards:
consolidating inefficient smallholdings, creating the motivation for investments
in productivity, and so on. The magnitude of these economic gains is a subject
of debate \citeeg{mccloskey72}. It is inarguable that commoners' lives were
upended by enclosure, but any honest accounting must acknowledge that their
lives before enclosure were difficult and desperately poor. Communities
sometimes consented to enclosure in the hopes of reaping its promised economic
rewards, which sometimes indeed materialized.

This article nonetheless casts its lot with a long interpretive tradition that
acknowledges the complexities of the enclosure process but believes that there
is an important conclusion to be drawn from the historical record. That
conclusion is: enclosure was a ``case of class robbery,'' indeed a ``revolution
of the rich against the poor'' \citep{thompson63}. In a word, enclosure was
theft. No small number of contemporaries believed this, as suggested by a
fragment of an anonymous poem quoted by \citet{boyle03}:
\begin{quote}
\textit{The law locks up the man or woman\\
Who steals the goose from off the common\\
But leaves the greater villain loose\\
Who steals the common from off the goose.}
\end{quote}
Thompson's use of the word ``revolution,'' however, is not accidental. Enclosure
also represented an upending of the existing social order. The theft of the
commons as land could not occur without the destruction of the commons as a
social system. Enclosure accomplished more than a mere transfer of wealth: it
erased the commoners' whole way of life.

Although nomenclature may refer to ``the'' enclosure movement in the singular,
it is noteworthy that enclosure played out in many communities over many
generations. Enclosure was not a focused and individually-driven theft like the
contemporaneous dissolution of the monasteries (1536--1541, instigated by Henry
\textsc{VIII}; \citealt{clark21}). Rather, it represents a change in the
circumstances of English life that disrupted a prior stable equilibrium. From
the historical context, it is difficult to avoid concluding that the change was
the emergence of capitalism. The commons were a form of wealth that were seized
by those who had the means and, newly, the motive to do so.

\section{Modern Commons}
\label{s.modern}

\noindent Even though the prototypical English commons are gone, they live in on
the broader concept of ``a commons'' and related theory. This larger idea tends
to resist a single, universally-accepted definition \citep{lessig02, boyle03}. A
convenient expression is that commons ``are things that no one owns and are
shared by everyone'' \citep{bollier25}.

While commons in the traditional sense are still studied
\citeeg{ostrom90}, modern scholarship is often concerned with a non-traditional
form: \b{information commons}. In economic terms, information is unlike the
pastures of yore in that it is non-rivalrous (if I'm using it, that doesn't
prevent you from using it as well) and non-excludable (it is expensive to
prevent someone from accessing and using it). These characteristics make
information resistant to being subjected to a private-property regime, and
correspondingly make it amenable to accumulation in a commons.

The general category of a commons is useful despite the stark differences
between information and agricultural land. Of particular relevance to this
article is the observation that all kinds of commons must be defended from
\b{defectors}, who attempt to appropriate the wealth of the commons for their
own private gain, and \b{incompetents}, whose participation is a net negative
for the community \citep{benkler02}.

The concept of intellectual property (IP) and its associated legal framework
(copyright, licensing, etc.) is how our society attempts to integrate
information into the property regime. Over the course of the twentieth century,
the scope of human creations that were treated as IP expanded dramatically, a
phenomenon that has been called ``the second enclosure movement''
\citep{boyle03}. This expansion is at least partially analogous to the what
happened centuries ago: the channeling of resources with communal, or at least
ambiguous, ownership into a private property regime.

Unlike the first enclosure movement, however, it is less clear that the
expansion of IP can be said to have upended whole ways of life. It is at least
plausible that strengthened IP protections actually helped some modern commoners
--- people that we could anachronistically call ``creators'' --- pursue their
crafts. The founders of the Free Software movement furthermore had the brilliant
insight that IP restrictions also have a role to play in creating and defending
information commons as a whole. The GNU General Public License (GPL) allows
people many freedoms in their use of GPL'd software, but requires that any
derivative works also be licensed under the GPL. This ``virality'' property
prevents would-be defectors from detaching themselves from the Free Software
commons.

Other kinds of information commons, on the other hand, can be thought of as
living in the gaps between IP boundaries. Most importantly, it is appropriate
and useful to think of human culture in general --- the whole kit and kaboodle
--- as an information commons that has a porous interface with the world of
intellectual property. While specific pieces of culture are, technically, owned
by various legal entities (``Marvel Cinematic Universe IP''), the general
``cultural commons'' largely exists outside of the property system, although the
border is ambiguous (Marvel's lawyers will come after toy counterfeiters but
not, hopefully, a child who doodles The Hulk on a notepad).

The concept of a commons is useful in domains beyond information and culture.
For an example particularly relevant to this article's intended audience,
consider the dark night sky. It is a resource of value shared by a number of
people (everyone who looks up) outside of the private property system. Like the
other commons discussed in this article, it is under threat \citeeg{lbr22}.

\section{Commons-Based Peer Production}
\label{s.peer}

\noindent Some information commons appear to arise spontaneously: a fascinating
book by \citet{sperber96} argues persuasively for an epidemiological
understanding of how the general cultural commons comes to be. But other
information commons are the result of more intentional effort. This section
summarizes the theory of \b{commons-based peer production} \citep{benkler02},
which offers an explanation for the origins of many modern information commons.

Benkler sought to understand the success of the open-source and Free Software
movement: how did an uncoordinated bunch of volunteer programmers on the
Internet manage to create an operating system (Linux) that could reasonably have
been said to be of higher technical quality than one produced by a
multi-billion-dollar corporation (Microsoft's Windows)? His answer turns out to
be relevant to science as well as software.

Benkler approached the question from an economic perspective. One can think of
software and scientific knowledge as goods that society produces, and consider
the processes by which this production occurs and is organized. Modern modes of
organization are typically stylized as markets and firms \citep{coase37}.
Benkler's core conclusion is that certain kinds of information goods --- ones
that require many small accumulations of diverse, specialized expertise to
produce --- are not just capable of being produced by loosely-organized groups
of people tending to a commons; it is in fact \i{more efficient} to produce them
this way than in markets or firms (see \citealt{benkler02} for the detailed
argument). This phenomenon is ``peer production.'' The key example of peer
production in Benkler's analysis is, of course, the open-source software
community. Others include Wikipedia, Project Gutenberg, and early forms of what
we would now call social media.

Strikingly, peer production is often conducted by volunteers in their spare
time. \citet{moglen99} proclaims this to be ``Moglen's Metaphorical Corollary to
Faraday's Law'': ``if you wrap the Internet around every person on the planet
and spin the planet, software flows in the network.'' It is a nigh-miraculous
empirical fact that if the conditions for peer production are present, people
need only the flimsiest of incentives to get to work.

Volunteerism is not, however, a necessary characteristic of the process. Indeed,
Benkler identifies ``the academic enterprise, and in particular science'' as the
``most important'' example of peer production. As in other peer-production
systems, in science the choice of who works on what task is largely decoupled
from both market pricing and top-down control, and the results of that work
accumulate in the broadly shared commons of the academic literature. The
importance of this commons can perhaps be seen in its negative: how outrageously
frustrating it is when a publication of interest is locked up behind a paywall.

On the other hand, one thing that is shared among all successful commons-based
peer production systems is the presence of some kind of \b{integration} process
(or processes), in which potential contributions are assessed and (broadly
speaking) either accepted into the commons or rejected. In the terminology of
the previous section, such a mechanism is needed to resist incompetents. In the
case of open-source software development, the most tangible example of
integration is the process of deciding to merge or reject a pull request. Peer
review is an obvious incarnation of the academic integration process.

The analysis in this section has been somewhat abstract, but it bears pausing to
reflect on some big ideas. First, commons-based peer production is a truly
amazing phenomenon. The story of the birth of Linux may be a bit stylized, but
its essence is true: a decentralized group of enthusiasts beat some of the
world's most powerful corporations at their own game --- a tangible example of
the capitalist production system being flatly outcompeted. Second, some of the
commons created by peer-production processes can fairly be said to represent
humanity's greatest achievements. Their value, both monetary and in the broader
sense, is beyond measure.

Finally: in this interpretation, scientists and scholars are commoners, in the
literal meaning that their way of life is bound to the commons, existing
somewhat outside of the market system. Unlike agricultural commoners, of course,
they generally enjoy high socio-economic status, an attribute they share with
the Luddites. Indeed, many of them consider the work they do, which creates and
sustains the commons, to be fulfilling and rewarding. (In the analysis of
\citet{arendt58}, they are truly performing work, unlike the agricultural
commoners who were laboring.) This is a feeling shared by many who contribute to
the commons nonprofessionally as well, a feeling that is all too rare in the
general economy. It is at least reasonable to argue that academic jobs and
culture are and remain distinctive thanks to their association with peer
production.

\section{AI and The Commons}
\label{s.ai}

\noindent At last, we turn to AI. Equipped with the historical context and
conceptual framework presented above, what conclusions can we draw?

\subsection{Large Language Models Could Not Exist Without the Commons}

\noindent To a good approximation, the major models like ChatGPT and Claude are
trained on the entire internet. Quantitative studies confirm the intuition that
the cost of creating a major model is dominated by the cost of the training
materials --- or, it would be, if the lion's share of the inputs were not
effectively free \citep{kr25}. Putting it another way, the training materials
are responsible for most of the value of the models; the work that actually
happens inside the offices of companies like OpenAI is, in comparison,
incidental. Another way to see this is that the different AI companies all reuse
the same fundamental training sets derived from public data (e.g., the Common
Crawl\fnurl{https://commoncrawl.org/}), even as they spend more and more money
to create proprietary training data. The analysis by Kandpal \& Raffel achieves
its result even with generous assumptions that underestimate the cost of
creating the Encyclopedia Brittanica by a factor of 400.

Discussion of the relationship between AI models and their training materials is
usually conducted through a legalistic lens, centering on questions of copyright
and IP law. The AI training sets indisputably include massive quantities of
copyrighted material \citep{gmsr24}. An OpenAI legal
filing\fnurl{https://www.telegraph.co.uk/business/2024/01/07/openai-warns-copyright-crackdown-could-doom-chatgpt/}
has stated that it ``would be impossible to train today's leading AI models
without using copyrighted materials.'' The legal implications of this are
currently being worked through in law journals, legislatures, and numerous cases
before the courts. Because the modern information commons are primarily managed
in an IP-oriented framework, and it is far from clear how the new phenomena of
AI training and AI-generated materials should be integrated into existing IP
law, the outcomes of these legal debates will have major implications for the
future of the AI industry. From the perspective of this article, however, the
issue is emphatically \i{not} how AI relates to copyright. That which is legal
may still be unjust.

\subsection{The AI Industry Exploits the Commons}

\noindent While their projections may be overly optimistic, the officers of the
major AI companies are at least hoping to earn revenues measurable in the
trillions of dollars. In comparison, the amount currently expected to flow back
into the hands of those who built the commons --- the millions of people whose
collective works are the true source of most of the models' value --- is a
pittance; currently in the low billions, taking that amount to be dominated by
the \apx\$1.5 billion \i{Bartz v.\ Anthropic} books
settlement\fnurl{https://www.npr.org/2025/09/05/nx-s1-5529404/anthropic-settlement-authors-copyright-ai}.
It is apparent that any financial returns for ``the commoners'' beyond token
amounts will have to be clawed out of these companies by force. In the
terminology of \autoref{s.modern}, the AI companies are incomparable defectors.

Exploitation of the information commons for profit is not a new phenomenon; this
could describe a movie based on the character of Sherlock Holmes. But only with
AI (specifically, LLMs) can this kind of exploitation operate at what one might
call ``industrial scale.'' Hand-in-hand with this scaling-up comes a loss of any
semblance of consent. While a movie producer might discuss their plans with the
estate of Sir Arthur Conan Doyle even if only as a courtesy, there's no way to
poll the collective authors of the entire internet and get permission for their
works to be used in AI training. From a legal standpoint, it may turn out that
consent is not necessary: use of many materials for AI training, even
copyrighted materials with restrictive licensing terms, may be judged to be a
form of fair use. If this comes to pass, it is certain that vast numbers of
people who have contributed to resources like Wikipedia will consider the
outcome unjust.

It can be argued that while the AI industry takes from the commons, it also
gives back. Generative AI applications create new material that can be
integrated into the commons, up to and including a staggering array of
cutting-edge mathematical results \citep{riehl26}. One could interpret AI
companies as analogous to fertilizer companies: people selling a useful product
that can provide an enormous benefit to productivity. Care would be advisable
before committing to this analogy \citeeg{sh18}. The essential distinction,
however, is that research and scholarship, like the subsistence farming of the
preindustrial commoners, are for the most part not profit-driven enterprises.
Meanwhile, the sellers of these products seek to amass fortunes.

\subsection{In Practice, AI Threatens Peer Production}

\noindent We are presently witnessing a dark version of ``Moglen's Metaphorical
Corollary to Faraday's Law,'' mentioned in \autoref{s.peer}. It is now clear
that if you give people generative AI tools and a network, they will spam slop.

Virtually every commons-based peer production process mentioned above is
currently facing the equivalent of a distributed denial-of-service (DDOS) attack
on its integration processes. In the arena of open-source software development,
this takes the form of rampant pull request and CVE
spam\fnurl{https://www.theregister.com/2026/02/03/github_kill_switch_pull_requests_ai/}.
Wikipedia has recently adopted a policy of prohibiting nearly all LLM-assisted
contributions since so many were
unacceptable\fnurl{https://en.wikipedia.org/wiki/Wikipedia:Writing_articles_with_large_language_models}.
The computer science section of arXiv had to adopt a new policy to deal with
numerous submissions of low-quality review
articles\fnurl{https://blog.arxiv.org/2025/10/31/attention-authors-updated-practice-for-review-articles-and-position-papers-in-arxiv-cs-category/},
and arXiv as a whole has imposed submission rate
limits\fnurl{https://blog.arxiv.org/2026/10/01/updated-rate-limit-policy/}.
More broadly, ``scholarslop'' is becoming prevalent, taking forms ranging from
hallucinated citations in broadly legitimate works \citep{ansari26} to more than
ten thousand untrustworthy biomedical articles published in 2025 alone
\citep{sohs+26}.

Returning to the vocabulary of \autoref{s.modern}, generative AI therefore
drives not just mass defection, but mass incompetence as well. This is not to
deny that AI is also fueling genuinely useful and even ground-breaking
contributions, as in the math results referenced above. But it is important to
understand that many of the people who are deeply involved in peer-production
integration mechanisms --- open-source maintainers, Wikipedia editors, journal
editors, and so on, who generally have strong ideological commitments to the
commons writ large --- view AI as the indisputable driver of unprecedented
difficulties. Besides the direct logistical impacts of slop spam, its
consequences for the motivation and enthusiasm of human participants in peer
production are dire.

\section{A New Kind of Enclosure}
\label{s.newenc}

\noindent While the expansion of IP in the twentieth century has been described
as a second enclosure movement, this article has argued that the analogy is
partial at best (\autoref{s.modern}). IP laws did broadly strive to turn
previously public property into private, but it is difficult to say that they
managed to overturn existing ways of life.

In terms of their fundamental mechanics, the changes driven by frontier AI are
also unlike the historical enclosure movement. Instead of seizing existing
resources, the AI industry deploys a new set of resources: trained models and
the associated infrastructure needed to make use of them. In a limited view, the
existing commons are left untouched.

But while the mechanics of the transgression may differ, its heart is the same.
Common wealth is being transformed into private gain at enormous scale. If the
concept of ``theft'' can be applied to anything broader than sample larceny, it
applies here. At least one Microsoft executive apparently agrees, having
described\fnurl{https://arstechnica.com/tech-policy/2026/09/microsoft-exec-called-ai-scraping-the-largest-theft-of-labor-in-human-history/}
the AI industry's actions as ``an astonishing theft of unprecedented
proportions,'' perhaps the ``largest theft of labor in human history.'' (Those
familiar with the slave trades might dispute this.)

Just as --- or perhaps even more --- significantly, unlike the growth of IP, it
is at least plausible that the growth of AI could permanently upend the social
structures that have evolved to grow and sustain the information commons. That
is, it does not seem unreasonable to believe that AI could fundamentally alter
the nature of the scientific and scholarly professions. If robots are writing
``papers'' (scare quotes because it seems implausible that they would stick to
existing forms or languages) at rates $10^5$--$10^9$ times faster than humans
\citep{dhogg26}, it seems that the commons must detach from direct human
management and comprehension altogether. In the field of software engineering,
the embrace of this vision has been described as a ``software factory'' and is
already being explored\fnurl{https://factory.strongdm.ai/}. It is hard to
envision what this would look like applied to the scientific commons, but at a
minimum it would represent a foundational change in what it means to do science
\citep[raising the question of why we do science;][]{dhogg26}. Even if one
prefers much more conservative projections, it is already possible to perceive
cracks in the foundations of the scientific and academic professions, as
discussed in the previous section.

In short, AI is precisely the tool needed to launch an unprecedented enclosure
of science, academia, and other information commons. It enables private actors
to appropriate the enormous wealth held in the commons at previously unthinkable
scales, and it simultaneously enables attacks upon the social systems that have
evolved to grow, nurture, and protect those commons.

Whether the transformed \i{feasibility} of enclosure will be followed by a new
enclosure \i{movement} is another question --- a fundamentally sociopolitical
one. But the basic economic incentives seem to point to an answer in the
affirmative. Some readers might argue that is what for-profit scholarly
publishers have been attempting, without the aid of AI, since Robert
Maxwell\footnote{Incidentally: the father of Ghislaine.} \citep{stevenson09}.
The cartoon version of the playbook is: first gain control of the crown jewels
--- the underlying literature --- then use that control as a basis to make
yourself indispensable to as many parts of the research process as possible,
aiming to own the entire ``research lifecycle'' \citeeg{moore25}. Historically,
the fundamental problem has been that you don't control the production of your
most precious asset, the literature. Your would-be customers, the researchers,
do, and they are an unruly bunch, a regrettable consequence of the fact that
research is most efficiently produced through peer production, not
command-and-control firms. Until now, the result has been something of a
stalemate.

This stalemate is being disrupted. Above, it was suggested that the first
enclosure movement arose because a new motive for theft (capital accumulation)
emerged to join pre-existing means (the power that the rich generically hold
over the poor). From this perspective, the information commons can be seen as
persisting simply because the means to appropriate their wealth \i{at scale} has
historically been lacking. A Sherlock Holmes movie here and there is one thing,
but how could one possibly turn the full breadth and depth of human culture into
private capital? This question now has a two-letter answer.

\section{Conclusion: Back to the Luddites}
\label{s.conc}

\noindent This article opened with reference to the Luddites, but has focused on
the commoners instead. This is for several reasons. First and foremost, the
author has observed that many working scientists are unfamiliar with the
relevant history and theory, and hopes to rectify that. In comparison, the story
of the Luddites may seem familiar. It can be cast as an early, clean example of
how conflict between labor and capital manifests as a struggle between human and
machine. To quote one of the all-time sloganeers in these matters: ``Machines
were, it may be said, the weapon employed by the capitalist to quell the revolt
of specialized labour'' \citep{marx1847}. This article will not attempt to
analyze AI from this viewpoint, but in this framework, scientists and scholars
are no different than workers in any number of other industries that depend upon
human expertise. The particulars of commons-based peer production are less
relevant. It is worth mentioning that in this interpretation the fundamental
distinction between skilled labor and capital is: who owns the machines?

Second, the Luddite perspective can (at least somewhat unfairly) be seen as
parochial --- the losers of economic history, complaining about being passed by.
It is to be hoped that in the public discourse, a concern for the integrity of
the commons, and resistance to attempts to steer them into private hands, help
convey that more than mere self-interest and resistance to change are at stake.

Finally, while conversations about AI can often fixate on the details and
capabilities of the technology, as well as projections about how those might
evolve in the future, the perspective of enclosure helps underline that the most
urgent concerns and discussions about AI really have little to do with its
technical aspects. One can view the enclosure movement through a technological
lens, but it was at its heart a phenomenon of economics, politics, and power.
The same is true of AI. The historian of technology Melvin Kranzberg articulated
six ``laws of technology'' \citep{kranzberg86}, the first of which has the
following form:
\begin{quote}
\i{Technology is neither good nor bad; nor is it neutral.}
\end{quote}
Kranzberg's First Law of Technology pithily captures a profound truth. Specific
technologies do not have intrinsic moral content; but the ways that they are
\i{used} in society absolutely do. It is a category error (as well as a
significant generalization) to say that AI is ``good'' or ``bad.'' But the ways
in which people choose, or are coerced, to use and develop it are emphatically
not neutral.

\begin{acknowledgments}
  This article emerges from a presentation and discussion at the UCSD Software
  Citation Workshop, April 2026. The author acknowledges support for the 2026
  Software Citation Workshop from the NASA TWSC program under award number
  316634-00001.

  PKGW thanks Rob Goodman and the Software Citation Workshop participants for
  helpful discussions, and Rosalind Williams for everything else.

  For the record, this article was written without using generative AI tools.
  PKGW just likes em-dashes. Supporting research used Google and other search
  engines, which these days can reasonably be said to constitute ``using AI.''
\end{acknowledgments}

\bibliography{index}{}

\end{document}